# On the Evolution of Dark Matter Halo Properties Following Major and Minor Mergers

Peter Wu[1,3] and Shawn Zhang[2,3]
[1] *Lynbrook High School, San Jose, CA 95129*
[2] *Amador Valley High School, Pleasanton, CA 94566*
[3] *Physics Department, University of California, Santa Cruz, CA 95064*

**Abstract**

We conducted an analysis on dark matter halo properties following major and minor mergers to advance our understanding of halo evolution. In this work, we analyzed ~80,000 dark matter halos from the Bolshoi-Planck cosmological simulation and studied halo evolution during relaxation after major mergers, those in which the merging mass ratio $m/M > 0.3$. We then applied a Gaussian filter to the property evolutions and characterized peak distributions, frequencies, and variabilities for several halo properties, including centering, spin, shape (prolateness), scale radius, and virial ratio. However, there were also halos that experienced relaxation without the presence of major mergers. We hypothesized that this was due to minor mergers unrecorded by the simulation analysis. By using property peaks to create a novel merger detection algorithm, we attempted to find minor mergers and match them to the unaccounted relaxed halos. Not only did we find evidence that minor mergers were the causes, but we also found similarities between major and minor merger effects, showing the significance of minor mergers for future studies. Through our dark matter merger statistics, we expect our work to ultimately serve as vital parameters towards better understanding galaxy formation and evolution.

# 1 INTRODUCTION

In the modern standard Lambda Cold Dark Matter (ΛCDM) theory on structure formation of our Universe, it is said that galaxies form and evolve within dark matter halos and subhalos [4, 7]. By analyzing the evolving properties of the halos, we can therefore better understand the underlying astrophysics of galaxies in the context of an expanding universe. One of the most important recent developments has been the emergence of large-scale cosmological simulations to resolve dark matter halos, giving us measurements of halo properties at many distinct time steps. Although it is expected that halos should typically grow in mass with cosmic time, it was recently discovered that there are times where halos experience "relaxation", in which they lose mass over a period of time [1]. Only recently has there been a study that details the cause of relaxation after major mergers, i.e., where the merging halos have a mass ratio $m/M > 0.3$ [5]. According to this work, major mergers initially boost halo mass and cause the halo to become more prolate and less relaxed and have higher spin and lower NFW concentration. Then as the halo relaxes, high energy material from the recent merger gradually escapes and reduces the halo mass typically by $5-15\%$. Halos that experience a major merger around redshift $z = 0.4$ will generally reach a relative minimum mass at $z = 0$. This same study, however, also noted that there were some halos that experienced relaxation-based mass loss without the presence of major mergers. Here we investigate whether such incidents may be caused by minor mergers. Although minor mergers (i.e., where the mass ratio $m/M < 0.3$) have not been as intensely studied as major mergers, they still have noticeable effects on halo properties [9]. By studying both major and minor mergers together, we are able to gain a more complete view of dark matter halo growth histories.

This paper's focus is to study the effects of mergers on dark matter halo properties, specifically in the context of relaxation. As of now, the fluctuation of halo properties post-merger has not been well-understood. Therefore, we provided here a comprehensive analysis of the distribution, frequency, and variability of the peaks in halo properties due to a major merger.

In addition, we also investigated the effects of "hidden" minor mergers. Since the main analysis of the Bolshoi-Planck cosmological simulation does not record minor mergers, we had to develop our own novel algorithm that detects merger events by property fluctuations. This is the first attempt to analyze such minor mergers, opening the door to re-evaluate past studies and conduct future ones. We then performed the same analysis we did on major mergers to minor mergers and found similar results. Ultimately, this work argues for the significance of minor mergers and the need to account for their effects in other dark matter halo-related studies.

# 2 METHODOLOGY

## 2.1 Data Overview

Our data came from outputs produced by ROCKSTAR (Robust Overdensity Calculation us- ing K-Space Topologically Adaptive Refinement), an algorithm for identifying dark matter ha- los, structure, and tidal features [2]. ROCKSTAR was run on the Bolshoi-Planck simulation, based on the Planck cosmological parameters [5, 6, 8]. Resultantly, we gained immense catalogs of ~80,000 halos with all their properties recorded at different scale factors $a = 1/(1+z)$, including when they experienced a major merger.

For our analysis, we focused only on halos within the mass range $10^{11.575} - 10^{12.325} M_\odot/h$.

Additionally, we kept our analysis limited to halos in low density environments, avoiding strong tidal effects and possible stripping. We also restricted our data to only after scale factor $a = 0.25$, i.e. $z \leq 3$, due to the low resolution at early times. Note: $a = 1$ ($z = 0$) represents the present time.

## 2.2 Halo Property Evolution

We mapped dark matter halos by their merger tree root (descendant halo at the last simulation snapshot) to distinguish them from other subhalos and recorded their properties chronologically. We could then plot the evolution of these halo properties for each individual halo. A sample halo's evolution is provided in Figure 1.

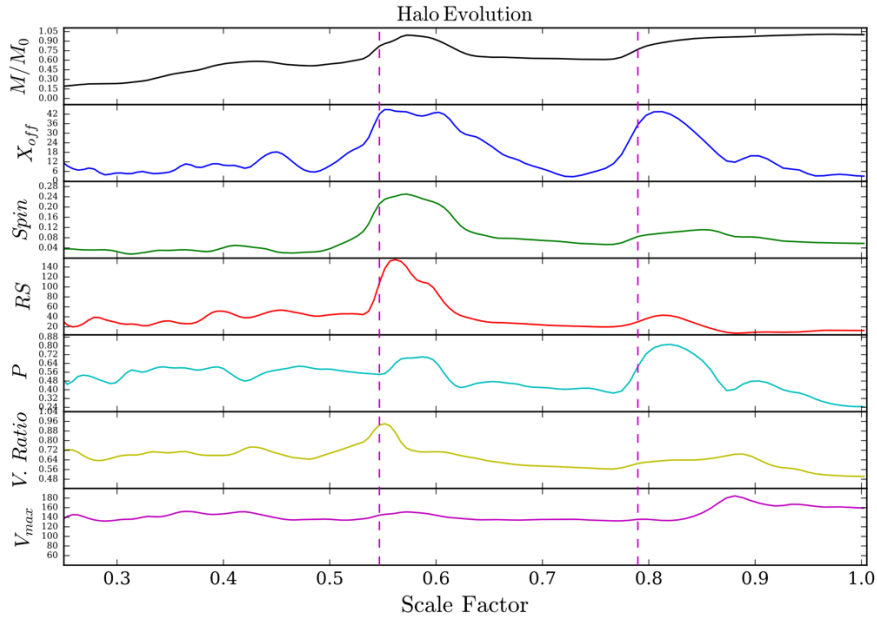

**Figure 1: Halo Property Evolution.** Seven properties are plotted on the y-axis with time in scale factor on the x-axis. The curves are slightly smoothed by a Gaussian filter. Mass $M$ is normalized by the halo's latest mass value, $M_0$. A dashed vertical line represents where a major merger event occurred.

We summarize the definitions of the halo properties that we looked at here: $M$ is the mass in units of $M_\odot/h$. $X_{off}$ is the offset of the halo center from the center of mass within the radius. *Spin* is the dimensionless measure of amount of rotation [3]. *RS* is the scale radius in kpc/$h$ where the logarithmic slope of the density profile equals -2. $P$ is the prolateness measure of halo shape from 0 (spherical) to 1 (needle shaped). *V. Ratio* is the virial ratio; ratio of halo kinetic to potential energy (0.5 for relaxed halos). $V_{max}$ is the maximum circular velocity in km/s.

## 2.3 Looking at Major Merger Distributions

In this study, we focused on five properties: $X_{off}$, spin, scale radius, virial ratio, and prolateness. Here, we detail our method to find where peaks typically occur in relaxation, how narrow the distributions are, and what the relative frequencies are among different mass loss bins.

### 2.3.1 Separating Halos by Mass Loss

To begin, we separated the halos into three bins: ones that have lost < 5%, 5−15%, and > 15% of their mass since their last major merger. This was done by finding the halo's peak mass value and comparing it to the current mass, $M_0$. Through the calculated percentage difference, we were then able to bin the halos.

### 2.3.2 Determining Typical Peak Locations During Relaxation

Next, we applied the following method to characterize how responses in the relaxation period differed among various mass bins. We first interpolated the evolution of the five properties (Section 2.3) onto an evenly spaced grid where the scale factors increment by 0.005 to set equal time intervals. We then applied a Gaussian filter at a standard deviation of $\sigma = 4.0$ (determined by various trials) to reduce the noise in our data. Because we wanted to concentrate on the relaxation period, we limited our window of peak detection so that the last major merger was at $0.45 \leq a \leq 1.00$ for each halo. This was performed to each mass bin along with a group containing all the halos.

Using $X_{off}$ as an example, we examined the smoothed data for regions throughout the $X_{off}$ evolution where $dX_{off}/dt = 0$ and $dX^2_{off}/dt^2 < 0$ (a peak) without detecting much noise. For the first peak after a major merger, we tabulated the scale factor change before it was detected. After doing this for all halos, we plotted the distribution of scale factors that we expected to elapse before the detected first peak for each mass loss bin. This is then applied analogously for detecting a second, third, and fourth peak. Furthermore, we reported the relative percentage of halos that peaked for each group by counting the total number of detected peaks within this window for each halo. By doing this for all halos, we calculated the relative percentages of peaks in each halo groups.

## 2.4 New Algorithm to Detect Minor Mergers

In order to gain a more comprehensive view on these halo properties after merger events, we needed to not only consider major mergers but minor mergers as well. Since ROCKSTAR does not record minor merger events, we had to figure out a way to detect them ourselves. Here, we describe a new simple algorithm to detect minor/major mergers. Note that although this algorithm analyzed the entire halo dataset, we display the following procedure through one specific halo.

### 2.4.1 Finding Property Fractional Differences

This algorithm centered around detecting common peaks among different halo properties. Because we wished to later correlate the five properties from Section 2.3 with our algorithm-detected mergers, it was important for us to use other properties and avoid circularity. Hence, we restricted ourselves to only use $M$ and $V_{max}$. For each of these two properties, we sought out fractional in- creases/decreases from one scale factor to the next using an interpolated 0.01 fixed step-size array. We solved this equation:

$$\delta_{prop}(a) = \frac{y_{prop}(a + 0.005) - y_{prop}(a)}{y_{prop}(a)} \qquad (1)$$

where $a$ is the scale factor, $y_{prop}$ is the value of the specific property at $a$, and $\delta_{prop}$ is the fractional change at $a$.

To continue, we needed to calculate the weighting for each property since they fluctuate in different rates and amounts. Separately for $M$ and $V_{max}$, we computed the average and standard deviation of the fractional change for (1) many randomly chosen scale factors from randomly chosen halos and (2) at the location of each known major merger. This then helped us correct for average time dependence and dispersion of each property and compare how strongly instantaneous changes in each property correlated with major merger events. Finally, the weighting was calculated by dividing the difference between both average fractional changes by the standard deviation of the randoms:

$$w_{prop} = \frac{\bar{x}_{\delta_{prop},MM} - \bar{x}_{\delta_{prop},rand}}{\sigma_{\delta_{prop},rand}} \quad (2)$$

where $w_{prop}$ is the property's weight, $\bar{x}$ is an average of fractional changes, and $\sigma$ is a standard deviation of fractional changes.

Then, we defined a new quantity, $Q(a)$, as the linear combination of $M$ and $V_{max}$'s fractional change – each corrected for average, dispersion, and weighting proportional to how strongly they correlate with major mergers. In other words, we determined how far away from the (random) average change each property was at a given scale factor and weighted it by how strongly changes in that property correlated with major mergers. Formally, $Q(a)$ was calculated through this equation:

$$Q(a) = \sum \left( \frac{\delta_{prop}(a) - \bar{x}_{\delta_{prop},rand}}{\sigma_{\delta_{prop},rand}} \times \frac{w_{prop}}{\sum w_{prop}} \right) \quad (3)$$

### 2.4.2 Establishing A Threshold

With $Q(a)$ calculated, we were able to see halo evolutions in various peaks. In order to discern which peaks could be classified as major mergers, we established an arbitrary threshold, describing any peaks above the said threshold to be locations where major mergers occurred.

We then needed to optimize this threshold to maximize agreement and minimize disagreement. In this case, agreement was the percentage of ROCKSTAR's actual major mergers that were contained within the algorithm's detection band. Disagreement was the percentage of our algorithm's band detections that did not contain an actual major merger. Starting from $Q(a) = 0.0$, we per- formed a linear brute-force optimization technique that moved the threshold up and recorded the difference between agreement and disagreement. Wherever the difference was the highest, we set that $Q(a)$ value as the optimal threshold. Here are our end results: *Optimal Threshold*: ~0.84, *Agreement*: ~93.03%, and *Disagreement*: ~35.81%.

Take into account that "disagreement" implied that although no major mergers were recorded there, the halo still experienced significant merger-like effects. We speculated that this disagreement was actually an indicator of a minor merger or a collection of minor mergers simultaneously occurring. Also, realize that although this threshold was optimized for detecting *major* mergers, it was not necessarily the best for detecting *minor* ones. In that sense, we needed to lower our threshold. However, our methodology did not have the ability to discern a "true" threshold. Hence, we created four thresholds that were 100%, 75%, 50%, and 25% of the original optimal threshold. More will be discussed in Section 3.3.

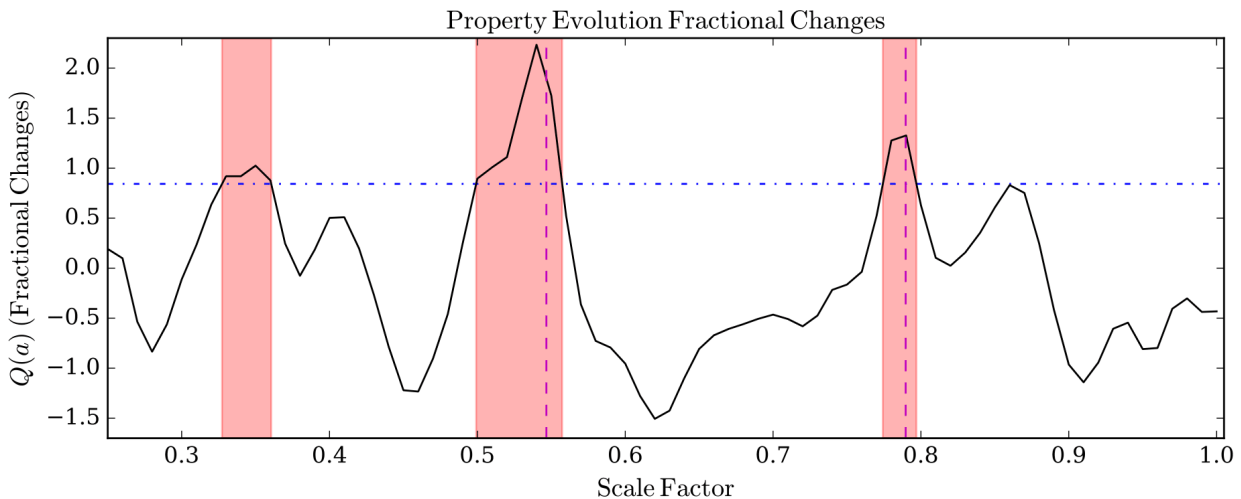

**Figure 2: Optimal Threshold Based On Fractional Changes.** This figure plots $Q(a)$ and displays the halo evolution in peaks. After establishing our optimized threshold (horizontal dashed line), we colored peaks that were above the threshold in red bands. These red bands are places where we determined merger effects occur.

### 2.4.3  Distinguishing Between Major and Minor Mergers

With our newly detected mergers, we needed to characterize between which events were major or minor. We wanted to distinguish between cases where the fractional change, $Q(a)$, just barely passed above the threshold and where it climbed higher. To do so, we solved the area between the peak and the threshold by taking an integral. Here is our equation:

$$A_T = \int_{a_{T_i}}^{a_{T_f}} (\delta_{overall} - T)\, da \qquad (4)$$

where $T$ is the threshold, $a_{T_i}$ is scale factor at the beginning of the band, $a_{T_f}$ is scale factor at the end of the band, and $A_T$ is the calculated area. Essentially, this gave us an idea about the total change in mass over the band we've identified. A greater change would indicate a higher mass ratio (major merger) and a smaller change would indicate a lower mass ratio (minor merger).

We then constructed a histogram of the halos for agreement and disagreement with the bins as $A_T$ in Figure 3. Evidently, disagreement seemed to dominate the lower bins. We then smoothed the histogram through kernel density estimation and found the intersection point (~0.07) between the agreement and disagreement curves. Mergers with an $A_T$ above the intersection point were declared major while others were declared minor.

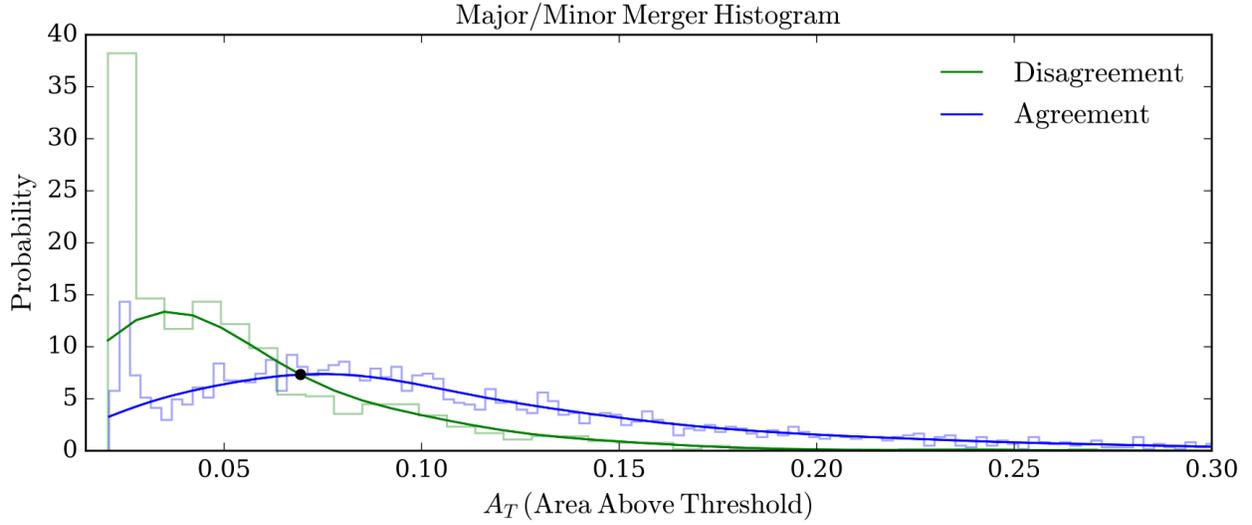

**Figure 3: Merger Classifier.** From the threshold, we plotted a histogram of the agreement and disagreement where the bins were based on Equation 4. By taking the intersection point between the agreement and disagreement curves, we made a classification point to characterize the mergers. Note: This plot is specifically of the 100% optimal threshold.

All in all, our simple algorithm systemically produced new outputs that for the first time, detail both major and minor mergers. As you can see in Figure 5, the algorithm's bands detected locations where merger events supposedly occurred. Both ROCKSTAR major mergers were contained albeit one of them was classified as minor. A supposed minor merger at $a$ ~0.5 was also discovered. This algorithm may not be perfect, but it proved to be adequate enough to reveal the significant effects of minor mergers. See Section 3.2 and 3.3.

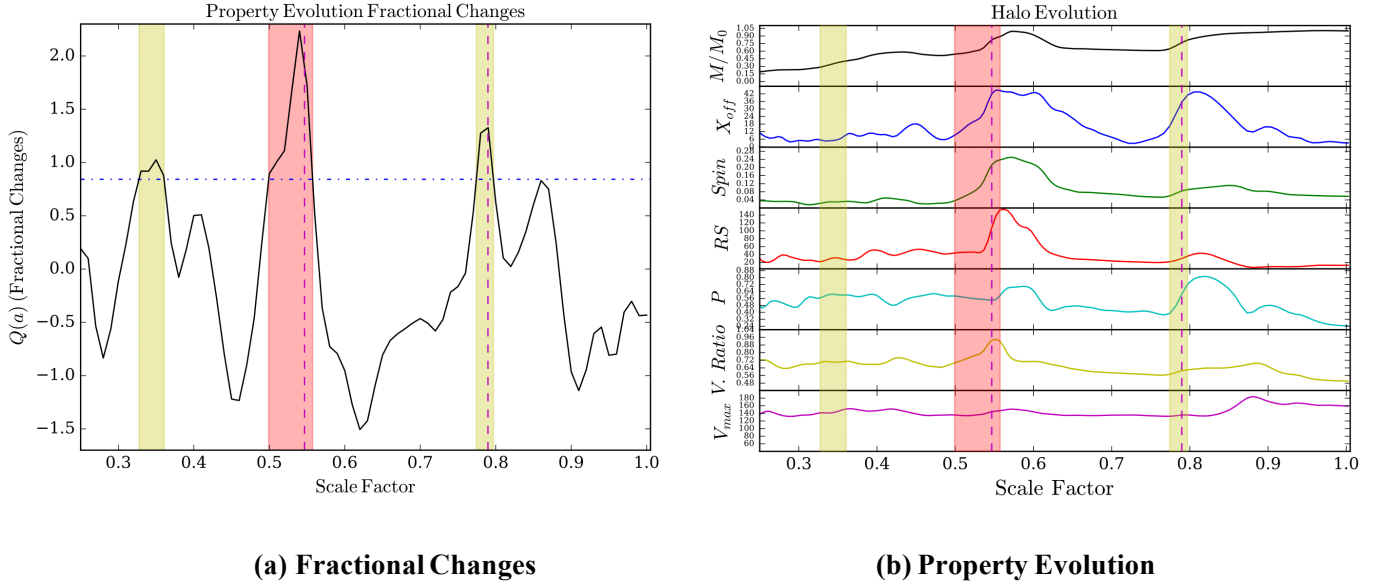

**(a) Fractional Changes**          **(b) Property Evolution**

**Figure 4: Major and Minor Merger Identification.** Based off the intersection from Figure 3, we were now able to differentiate between major and minor mergers. From Figure 1 and Figure 2, the bands now labeled merger events. A red band signifies a major merger and a yellow band signifies a minor merger.

## 2.5 Looking at Minor Merger Distributions

We applied the same methodology as in Section 2.2, but instead of tracking halo properties after a major merger, we tracked them after the predicted minor mergers using the algorithm in Section 2.3. After doing so, we compared property behaviors and tested the robustness of our algorithm.

## 3 RESULTS & DISCUSSION

### 3.1 Halo Properties During Relaxation Analysis

This section displays the analysis of peaks following a major merger in the five halo properties outlined in Section 2.2. The following subsections analyze each property in detail with a concluding section to synthesize the results, compare and contrast peak responses, and provide explanations.

#### 3.1.1 $X_{off}$

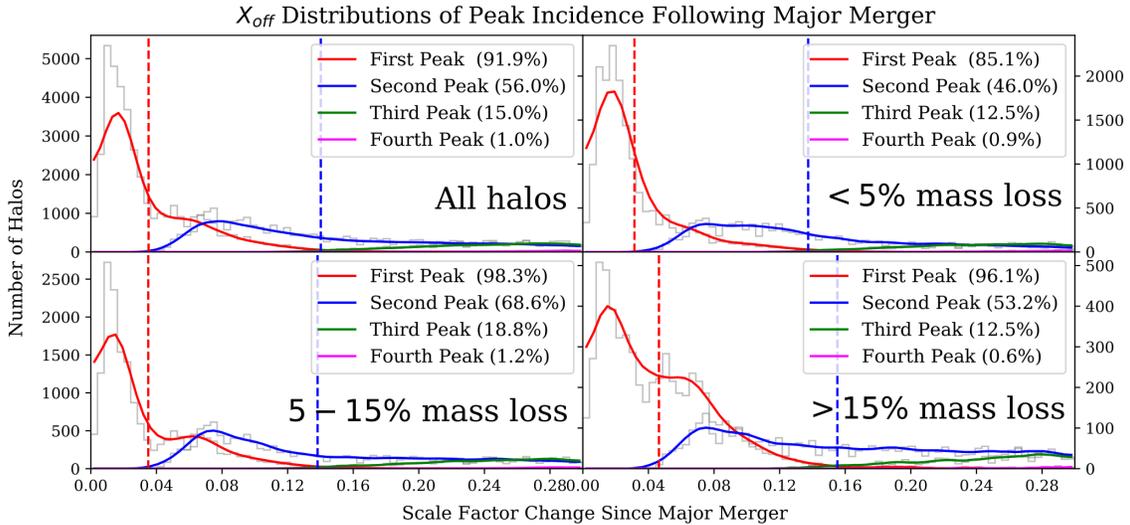

**Figure 5: $X_{off}$ Peak Distributions During Relaxation.** This plot shows four panels containing four groups of halos. In each group, a histogram is plotted for the distribution of scale factors that elapse before a first, second, third, or fourth peak is detected in $X_{off}$ – indicated by their respective color. The legend reports the relative frequency of halos within each particular group that exhibited a particular number of peaks. In each panel, the red dashed line indicates the average scale factor that elapsed before the first peak was detected within each group, while the blue is represented for the second peak.

During relaxation, the first peak in $X_{off}$ was almost always exhibited by halos (~0.04 scale factor change after the major merger) and a second peak (~0.13 scale factor change) was detected about 56% of the time. Third and fourth peaks were not particularly common, and if occurred, happened long after the major merger. Halos that had lost $5-15\%$ of their mass displayed the strongest peak tendencies, as among these halos, nearly 69% of them displayed a second peak. The appearance of a second peak was stronger in the $> 15\%$ mass loss bin than the $< 5\%$ mass loss bin by roughly 7%. The narrow distribution of the scale factor changes for the second peak in $X_{off}$ suggests that this was caused by the major merger directly.

### 3.1.2 Spin Parameter

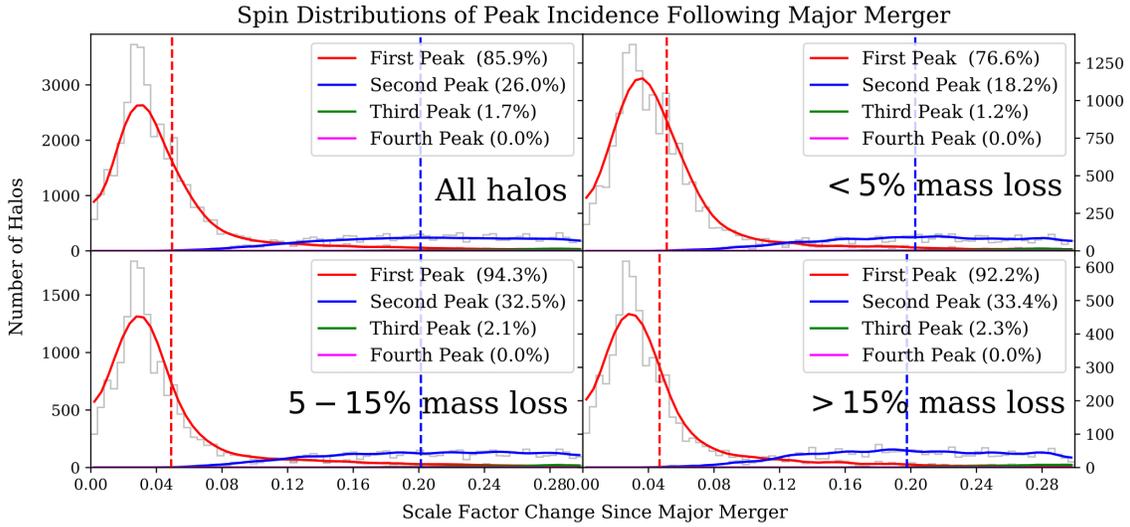

**Figure 6: Spin Peak Distributions During Relaxation.** Analogous to Figure 5 but examines the spin parameter.

The spin parameter exhibited slightly different results from $X_{off}$. They were similar in that the first peak was detected most of the time, as nearly 86% of halos had a first peak at ~0.04 scale factor after the major merger. The difference was in the less frequent second peak, as it was only detected in 26% of halos at a farther time step ($a \approx 0.20$). The second peak was slightly stronger in the halos that had lost $> 15\%$ mass loss than those that had lost $5-15\%$. The broad distribution of the second peak implied that second peaks in spin were likely not caused by the merger itself but rather by extraneous factors. Third and fourth peaks were even weaker and farther from the major merger in spin than they were in $X_{off}$.

### 3.1.3 Scale Radius

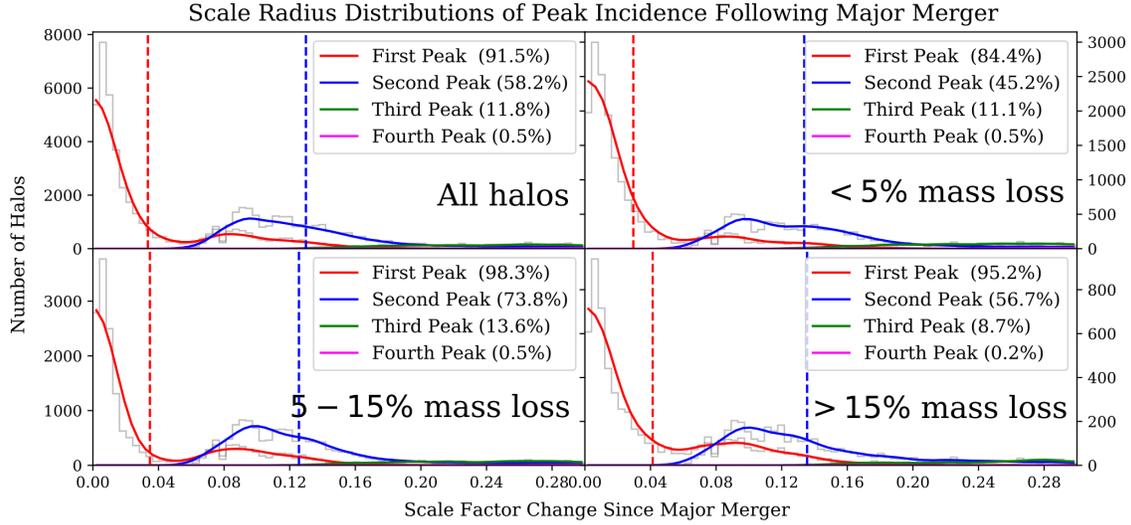

**Figure 7: Scale Radius Peak Distributions During Relaxation.** Analogous to Figure 5 but examines the scale radius.

Scale radius was similar to the previous two properties in that a first peak was usually always detected (~0.04 scale factor change) and a second peak was frequent (even stronger than $X_{off}$). The $5-15\%$ mass loss bin displayed the strongest peak tendencies among all peaks, especially the second peak as almost 74% of halos within this mass bin had a second peak (~0.12 scale factor change). The narrower distribution of the second peak implied that scale radius does seem to peak directly due to the merger. The third and fourth peaks in scale radius are closer to the major merger and more frequent than in spin parameter; however, they were still relatively weak when compared to the frequencies of the first and second peaks.

### 3.1.4 Prolateness

Prolateness was similar to the other properties with its first peak detected in almost 90% of the halos at the same time frame of ~0.04 scale factor change. Meanwhile, a second peak was observed roughly half the time. We see that the $5-15\%$ mass bin continued to display the strongest peak tendencies in the second peak as almost 58% of halos in this group had a second peak (at $a \approx 0.12$). Like in $X_{off}$ and scale radius, the narrow distribution of the second peak indicated that prolateness was likely caused by the major merger itself. Third and fourth peaks continued to be weak and not common among the halos.

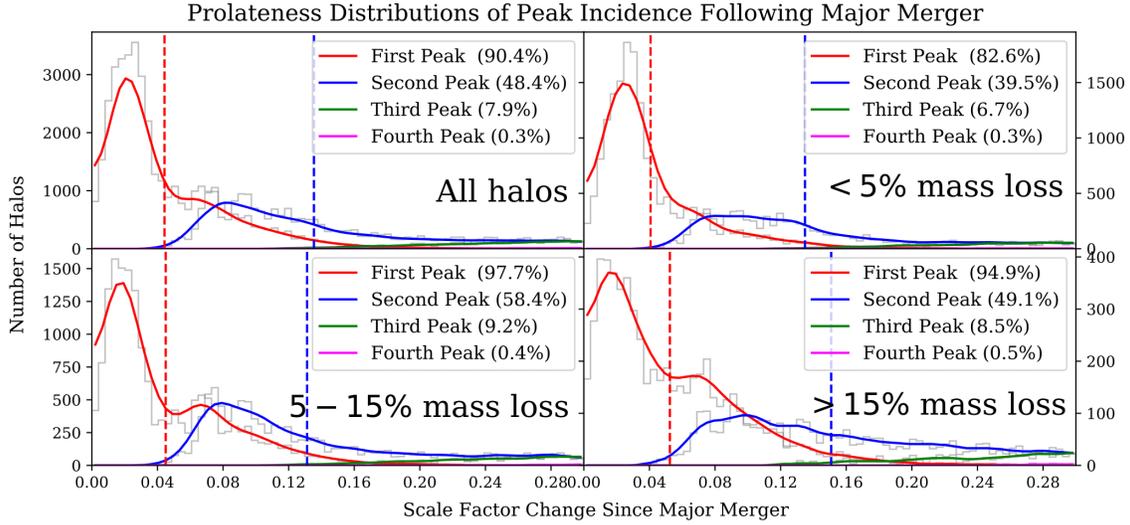

**Figure 8: Prolateness Peak Distributions During Relaxation.** Analogous to Figure 5 but examines the prolateness.

### 3.1.5 Virial Ratio

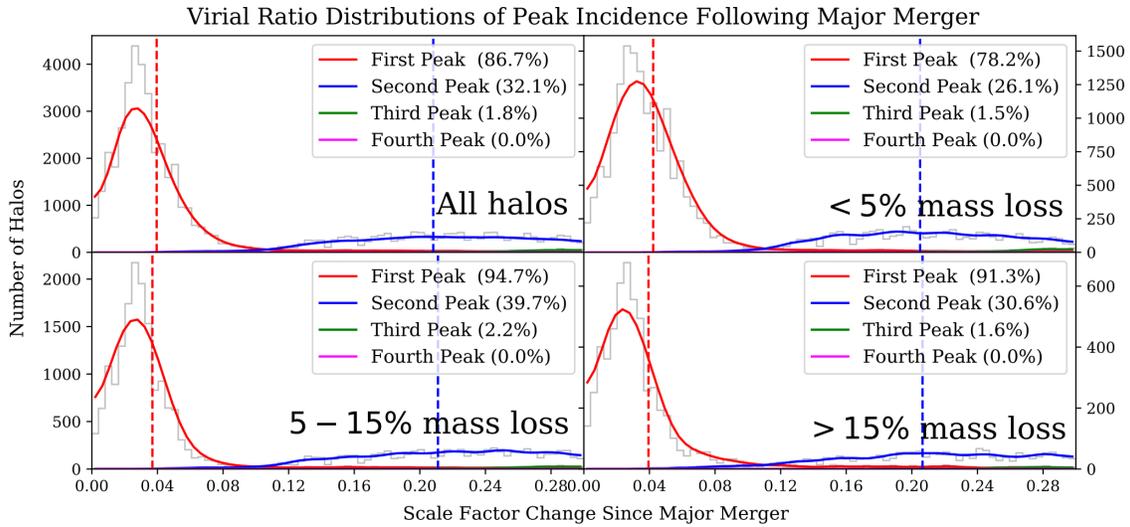

**Figure 9: Virial Ratio Peak Distributions During Relaxation.** Analogous to Figure 5 but examines the virial ratio.

The last property we observed was virial ratio which appeared to be very similar in peak response with the spin parameter. As was the case in the other four properties, the first peak was strong (87% of all halos) and detectable immediately (~0.04 scale factor change) after the major merger. The second peak was not frequent (only about 32% of halos) and far ($a \approx 0.20$) after the merger. Like spin, the distribution of the second peak was broad and widely distributed which suggested that the second peak was likely not caused by the merger itself. The 5 −15% mass bin continued to exhibit the strongest peak tendencies across all peaks in relative percentages. The third and fourth peaks continued to be almost nonexistent during the relaxation period halos experience.

### 3.1.6 Trends in Halo Properties After Major Mergers

From Figures 5-9, we were able to draw a multitude of conclusions about the peaks in halo properties during relaxation. First, nearly all major mergers caused a peak immediately for these five properties with a general delay of $a \approx 0.4$ or ~9.5 gigayears. This prominent commonality could allude to a tighter connection between halo properties in the context of merger events.

The second peak revealed even more interesting results. $X_{off}$ and scale radius typically exhibited two peaks in the 5 −15% mass loss group, while prolateness appeared to exhibit a second peak about half the time. Furthermore, this second peak was found closer in time around $a = 0.13$. However, the second peak for spin and virial ratio happened much later after the merger (~0.21 scale factor) and much less frequently.

As stated in [5], the physical properties of halos begin to change dramatically when an incoming halo impinges upon the host halo, becomes tidally disrupted, and deposits material onto the host halo. This is due to the dense core of the merging halo punching through near the center of the host halo and splashing back. We believe that the presence and interaction of
these two cores would tend to strongly influence properties like $X_{off}$, scale radius, and prolateness, as opposed to properties like spin parameter and virial ratio (which increase initially but do not respond to splash-back effects). Thereafter, the core of the merging halo becomes fully disrupted and is integrated into the host halo, settling back to lower values of scale radius, spin, prolateness, and virial ratio. This very likely explains why a third or fourth peak was not particularly strong, as even if it was detected, it was not common and happened at a much later time since the major merger.

Additionally, it was observed that dark matter halos that had lost 5 −15% of their mass since their last major merger were more prominent than others. As this was the most common level of induced mass loss, it came as no surprise that this particular group of halos had an overall highest tendency to exhibit one or more peaks for all five properties. The > 15% mass loss bin did not show strong peak tendencies across the properties likely because major mergers do not commonly produce such a severity of mass loss. Halos that lose this much mass appear to be indicative of outliers that lack a clean merger or reveal factors other than relaxation.

Overall, these results reinforce [5] and provide further characterization for dark matter halo properties.

## 3.2 HALO PROPERTIES AFTER MINOR MERGER

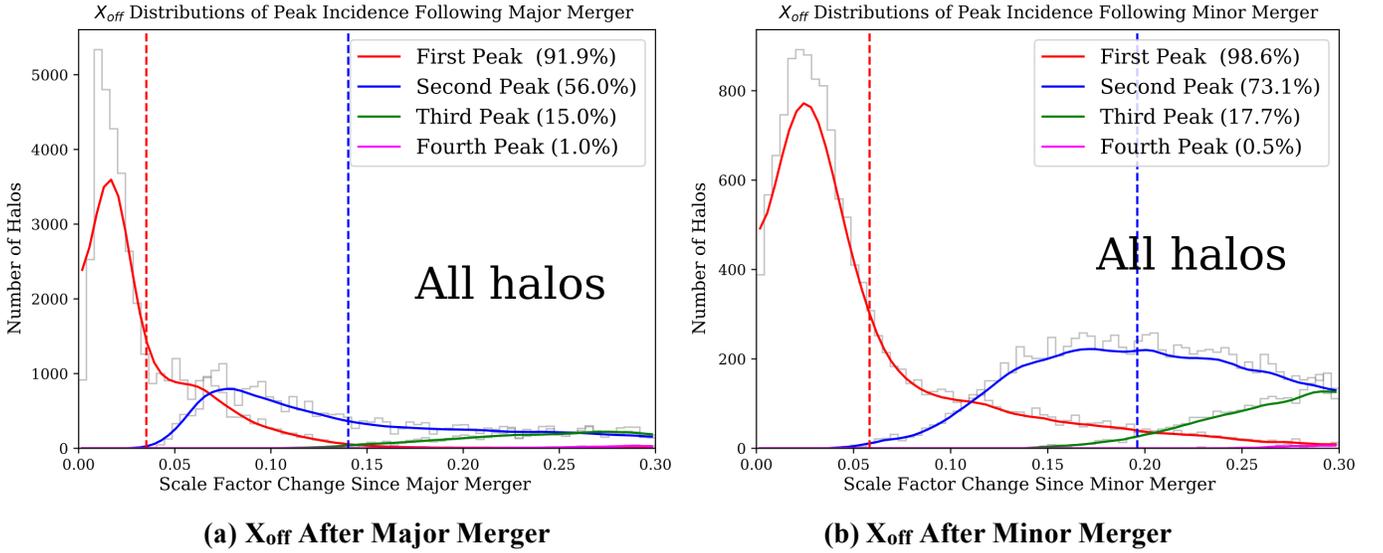

(a) $X_{off}$ After Major Merger

(b) $X_{off}$ After Minor Merger

**Figure 10: $X_{off}$ After Major and Predicted Minor Mergers.** The left panel displays the peak incidence distribution of $X_{off}$ after a major merger, while the right panel displays that for $X_{off}$ after a predicted minor merger using our algorithm described in Section 2.4.

Using the same methodology of studying distributions of peak incidence as we did in Section 3.1, we applied this to the predicted minor mergers we obtained from Section 2.4 (at 75% optimal threshold). By examining Figure 10, we concluded that $X_{off}$ behaved very similarly in peak distributions for both kinds of merger events. Both almost always caused a peak in $X_{off}$ immediately after the merger. The narrow distribution of the second peak for minor mergers suggested that this second peak in $X_{off}$ was directly caused by the minor merger.

We compared the major and minor merger responses from the other four properties. From Figure 11, we could see that a minor merger caused a peak in spin, scale radius, prolateness, and virial ratio following the merger. The spin and virial ratio for the first peak were the narrowest, while scale radius and prolateness had a delayed and broader range of peak times. Scale radius and prolateness distributions for the second peak were weakly peaked, suggesting that some of it may be due to the minor merger. The broad distributions of the second peak for spin and virial ratio suggested that this was generally not caused by the merger, a similar result observed from major merger responses described in Section 3.1.6.

The generally higher second and third peak rates for minor mergers compared to major mergers revealed interesting results. Although we expected that multiple peaks rates in scale radius, prolateness, and $X_{off}$ would be higher following major mergers than minor mergers, this could reflect the significance of the first peak in these properties. For a major merger, these quantities could dramatically increase (first peak) and could take a while to settle back down, *suppressing* the occurrence of subsequent non-merger induced peaks. If minor mergers only increase these quantities weakly (first peak), then it is likely that subsequent non-merger induced fluctuations may show up sooner and more often, accounting for this discrepancy. Future studies may have to pay closer attention on evaluating these first peaks in halo properties.

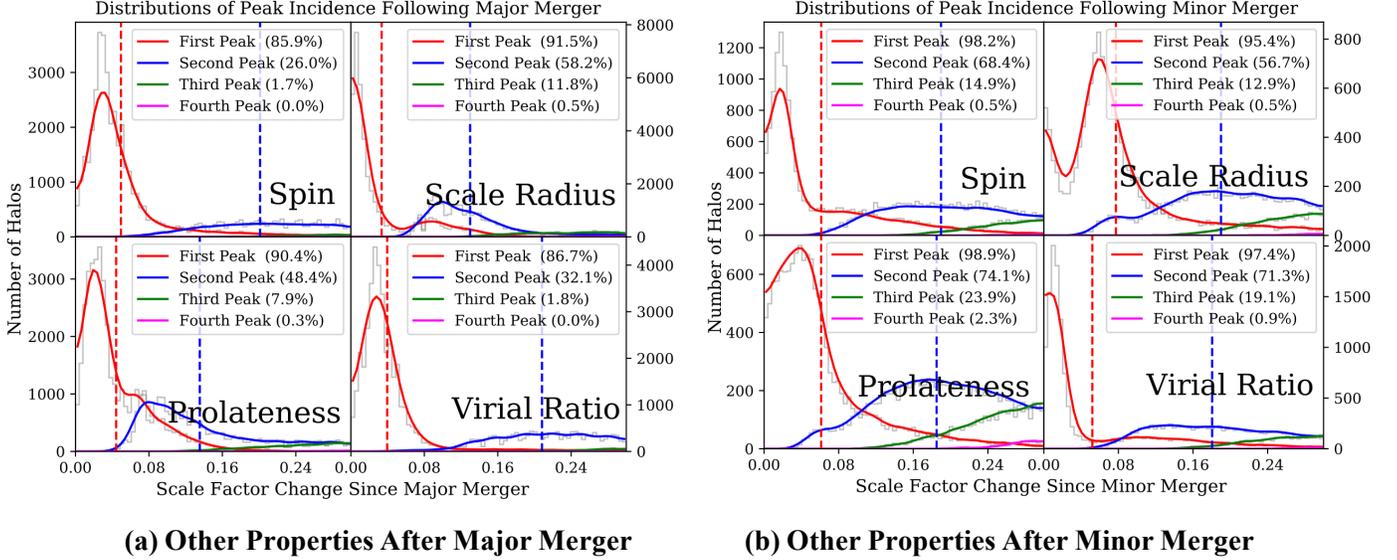

**(a) Other Properties After Major Merger**   **(b) Other Properties After Minor Merger**

Figure 11: **Spin, Scale Radius, Prolateness, And Virial Ratio Peak After Major and Predicted Minor Mergers.** Analogous to Figure 10, this figure displays major merger peak incidence distributions for the other properties (spin, scale radius, prolateness, and virial ratio) after a major merger on the left, and for predicted minor mergers similarly on the right.

As a whole, our various distributions provided evidence that minor mergers act extremely similar to major mergers – both greatly influence property behaviors. Although they were "hidden" before, minor mergers can now be used to identify unexplainable property peaks. We emphasize that it is crucial to also account for minor merger effects in future dark matter halo studies.

### 3.3 Determining the Source of Unaccounted Relaxation

Stated before in Section 1.1, although [5] concluded that halo relaxation was caused by major mergers, there were quite a few exceptions where halos lost mass without any noted causes. Seeing the prominent effects from Section 3.2, we hypothesized that this unac- counted relaxation-based mass loss was due to minor mergers unrecorded by ROCKSTAR. With our algorithm from Section 2.4, we were now able to test our hypothesis.

We first found a subset of halos that experienced > 5% mass loss without any recent major mergers. Then using the merger detection algorithm, we found which of such halos also had a predicted minor merger from $a = 0.45 - 1$. Now, remember from Section 2.4.3 that the algorithm's optimal threshold was not well-designed to detect all minor mergers. Thus, we conducted our analysis in 4 groups: 100%, 75%, 50%, and 25% of the original optimal threshold. Running through this data subset, we found a dominance of predicted minor mergers (as long as we decreased the threshold) with very few predicted major mergers. When we compared the distribution of minor merger-driven mass loss from this subset with the distribution of major merger-driven mass loss from the entire dataset, we found striking similarities in shape, center, and spread. Please look to Figure 12.

The similarities between these two histograms led us to believe that minor mergers did indeed cause post-merger relaxation mass loss. Just like major mergers (see Section 1.1), minor merger events around $z = 0.4$ will cause a relative minimum mass at $z = 0$. Hence, that is the reason why both distributions are centered around $a = 0.7$. Once again, we demonstrated that minor merger effects significantly affect dark matter halo evolutions and could be the causes for inexplicable property behaviors. These agreeable results between major and minor mergers may hint for a necessary re-evaluation of the 0.3 mass ratio merger classification.

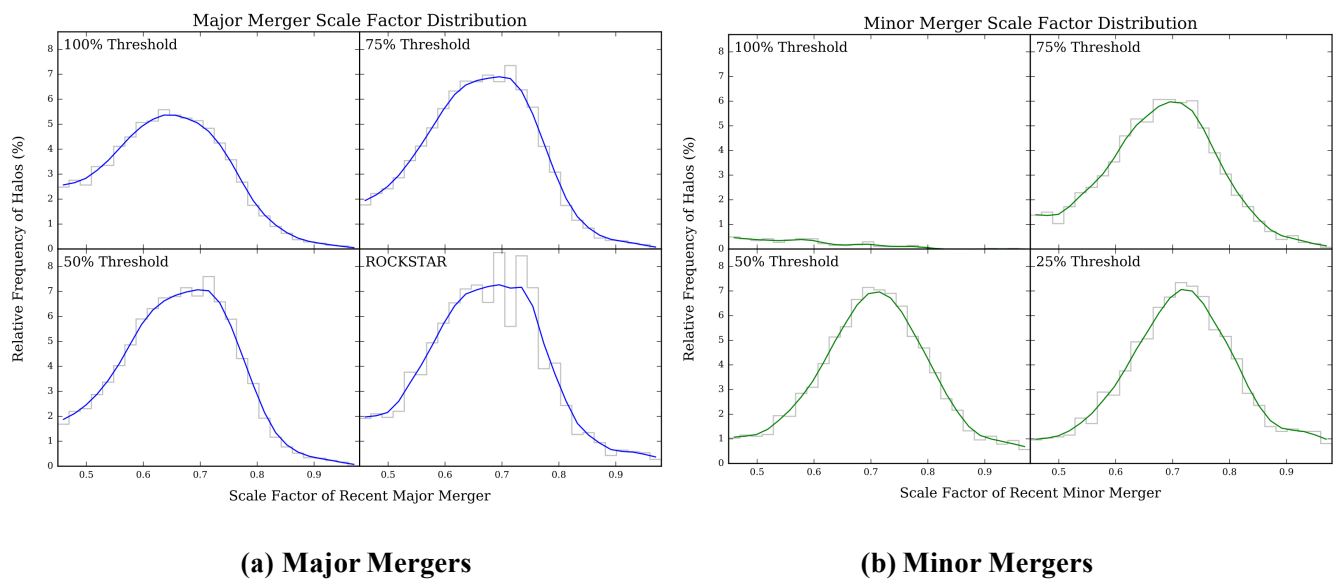

(a) **Major Mergers**  (b) **Minor Mergers**

**Figure 12: Major And Minor Merger Comparison.** The left subfigure plots the distribution of major merger-driven mass loss from the entire halo dataset. The right subfigure plots the distribution of minor merger-driven mass loss from a subset of data. Notice the similarities in shape, center, and spread. Each panel's merger events were obtained by the method located in the upper-left corner.

Note: All these agreeable results inform us that our algorithm had worked rather thoroughly. However, just as assumed, our original (100%) optimal threshold was too high and barely detected any minor mergers. However, when we lowered the threshold slightly, we found a noticeable increase in detections. This may provide us some clue for finding the "true" threshold.

# 4 CONCLUSIONS

In this study, we were able to characterize the distribution of responses for various dark matter halo properties following a major merger ($X_{off}$, spin parameter, scale radius, prolateness, and virial ratio), including the degree to which some properties could peak multiple times. Figures 5-9 provide a comprehensive and coherent picture of specifically how, where, and why each halo property exhibits peaks during relaxation. By studying these dynamics, we found a potentially tighter connection between the halo properties and profoundly better characterized halo behaviors. Additionally, we provided a new simple yet efficient algorithm that expanded upon ROCKSTAR's halo finder and detected merger events through our work in property behaviors. This algorithm was the first of its kind to also record minor mergers, opening a new study of obscure minor merger events and giving it high potential to be employed in the future.

Using the algorithm, we applied the same analysis as we did for major mergers and applied it to our predicted minor mergers and found similar results. Not only that, but we provided evidence that these minor mergers were actually driving factors behind unaccounted relaxation-based mass loss. Although minor mergers are currently poorly understood, these results prove that future studies must explore the effects of minor mergers more in-depth to gain a more comprehensive view of dark matter evolutions. We even consider potentially changing the 0.3 mass ratio merger classification – at least in the context of relaxation. In the end, our project's analyses on halo properties following major and minor mergers should serve as vital parameters towards better understanding galaxy formation and evolution.

# 5 FUTURE WORK

In our future work, we intend to study correlations in other dark matter halo properties. We also may apply our peak analyses of halo properties on areas not directly after a merger event. If we are able to find anomalies in these properties not explained by the relaxation phenomena, we may be able to find unexplored effects in the formation and evolution of galaxies (similar to our work with minor mergers).

Furthermore, we hope to improve our merger detection algorithm. As of now, we are not able to decide on a "true" threshold. By combining fluctuations in property peaks and the frequencies of relaxation-based mass loss (like in Figure 12), we may be able to develop a method that classifies minor and major mergers with better accuracy. In addition, we wish to be able to characterize mass ratios between merging halos in order to quantify the expected results. By doing so, we just might be able to evaluate a better mass ratio to classify between major and minor mergers.

Most importantly, we aim to use our results to advocate for additional studies in minor mergers. We plan to contact and collaborate with the ROCKSTAR team and develop an algorithm that details minor mergers in a more extensive way. Correspondingly, we will be able to push our understanding of dark matter merger statistics to further levels.

**Acknowledgments:** We thank the Science Internship Program at UCSC, organized by Prof. Raja GuhaThakurta, within which this research was done, and we thank UCSC astrophysics graduate student Christoph Lee and Physics Prof. Joel Primack for supervising this research.